\documentclass[acmsmall, screen]{acmart} %
\AtBeginDocument{%
  \providecommand\BibTeX{{%
    \normalfont B\kern-0.5em{\scshape i\kern-0.25em b}\kern-0.8em\TeX}}}

\setcopyright{acmcopyright}
\copyrightyear{2024}
\acmYear{2024}
\acmJournal{TORS}
\acmVolume{1}
\acmNumber{1}
\acmArticle{1}
\acmMonth{1}
\acmDOI{10.1145/3700890}

\usepackage[normalem]{ulem}
\usepackage{todonotes}
\presetkeys%
    {todonotes}%
    {inline,backgroundcolor=yellow}{}

\begin{document}

\title{A Survey on Intent-aware Recommender Systems}

\author{Dietmar Jannach}
\email{dietmar.jannach@aau.at}
\affiliation{%
  \institution{University of Klagenfurt}
  \country{Austria}
}
\author{Markus Zanker}
\email{markus.zanker@unibz.it}
\affiliation{%
  \institution{Free University of Bozen-Bolzano and University of Klagenfurt}
  \country{Italy and Austria}
}

\begin{abstract}
Many modern online services feature personalized recommendations. A central challenge when providing such recommendations is that the reason \emph{why} an individual user accesses the service may change from visit to visit or even during an ongoing usage session. To be effective, a recommender system should therefore aim to take the users' probable \emph{intent} of using the service at a certain point in time into account. In recent years, researchers have thus started to address this challenge by incorporating \emph{intent-awareness} into recommender systems. Correspondingly, a number of technical approaches were put forward, including diversification techniques, intent prediction models or latent intent modeling approaches. In this paper, we survey and categorize existing approaches to building the next generation of \emph{Intent-Aware Recommender Systems} (IARS). Based on an analysis of current evaluation practices, we outline open gaps and possible future directions in this area, which in particular include the consideration of additional interaction signals and contextual information to further improve the effectiveness of such systems.

\end{abstract}

\ccsdesc[500]{Information systems~Recommender systems}

\keywords{Recommender Systems, Intent-Awareness, Survey}

\maketitle

\section{Introduction}
\label{sec:introduction}

Recommender systems play a vital role for many modern online services, e.g., in e-commerce or media streaming, where they can create substantial value both for consumers and service providers~\cite{Jannach2021RSHBImpact}. Correspondingly, a multitude of technical approaches to generate personalized recommendations have been proposed over the last decades~\cite{RSHandbook2022}. Traditional algorithms---beginning with the early GroupLens system~\cite{ResnickGrouplens1994}---usually consider the entire history of known user preferences to create recommendation lists. As a result, such recommendations typically reflect the potentially diverse set of user interests and needs that were observed over time.  However, in many cases when we access an online service, we do so with a particular goal or \emph{intent} in mind. For the case of music streaming services, for example, our goal might either be to quickly access our favorite music, to find music to play in the background, or to discover interesting new artists~\cite{mehrotra_jointly_2019,kapoor_i_2015}. Likewise, in an e-commerce setting, we might access an online shop to research the set of choices in a certain item category, to continue our previous shopping session, or just to browse the catalog for inspiration~\cite{shi_improving_2017}. As a result, the set of suitable recommendations can largely depend on the user's underlying intent, which may not only change from one visit of the service to the next but even during an ongoing usage session.

Given these diverse sets of potential short-term intents, a recommendation algorithm that considers all of our past preferences independently of the user's current situation and goals may thus not be optimal. Since it can be difficult to correctly guess a user's specific intent when accessing the service, a common solution \emph{in practice} is to provide users with several recommendation lists on the landing page of the service, each of them addressing a different potential intent. Multiple recommendation lists are commonly used by media streaming providers such as Netflix or Spotify and on e-commerce sites like Amazon.com~\citep{Gomez-Uribe:2015:NRS:2869770.2843948,mehrotra_jointly_2019,Loepp2023Multi}. Alternatively, some services try to diversify their single-list recommendations to cover multiple potential user intents. They may, for instance, comprise items that presumably constitute novel and serendipitous discoveries to a given user while, at the same time, a small number of items the user has recently already seen is also included in recommendation lists. Due to the latter the user could conveniently continue a previous shopping session~\cite{Lerche2016}.

In the academic literature, the problem of diverse and time-varying user needs has been identified and addressed in different ways over time. In the realm of the traditional way of building recommender systems based on a given user-item interaction matrix, common approaches include intentional diversification of recommendations, temporal information and interest drifts, or utilization of contextual information like the user's current location and time~\cite{DBLP:journals/umuai/CamposDC14,bogina2022considering,Adomavicius2015,Kaminskas:2016:DSN:3028254.2926720}. More recently, academic research has largely shifted from the traditional ``matrix completion'' recommendation setting to \emph{sequence-aware} and session-based recommendation problems~\cite{QuadranaetalCSUR2018,wang2021survey}. In these approaches, generated recommendations are commonly determined by considering the most recent interactions observed for a given user. The main underlying assumption of these approaches is that these recent interactions are a good reflection the user's underlying short-term intents.

However, while the most recently observed user actions may be indicative of the underlying user intents, researchers have recently tried to incorporate intent-awareness into recommender systems in a more explicit manner. Importantly, reports from industrial applications emphasize the strong potential of adapting recommendations to the predicted intent(s) of users. For instance, a field test in the e-commerce domain~\cite{shi_improving_2017} revealed that relevant business KPIs (key performance indicators) like purchase rates can be markedly improved when the recommendation system dynamically switches its strategy depending on the currently assumed user intent. Likewise, a study in the music streaming domain by Spotify~\cite{mehrotra_jointly_2019} investigated the connections between observed user interactions, underlying intents and user satisfaction. One main conclusion of this work is that understanding user intents can be crucial to predict user satisfaction, and that different interaction signals are relevant as predictors, depending on the user's underlying intent.

Given the high practical relevance of such approaches, the aim of our present work is to review the growing body of literature on \emph{intent-aware} recommender systems (IARS). For this purpose, we have identified a larger set of recent works based on a semi-systematic literature search, and categorized these works along different dimensions. A particular focus of this survey lies on the technical approaches to incorporate intent-awareness into the system, such as diversification techniques, intent prediction models or latent intent modeling approaches. We furthermore review common application domains and evaluation procedures for IARS, which revealed a very strong focus on offline experimentation with an emphasis on prediction accuracy as the main metric. Finally, our survey helped us to identify open gaps in the literature and future research directions. Specifically, the increasing availability of additional pieces of information about users and their behavior as well as on item features and contextual factors opens continuously new opportunities to better predict a user's specific intent when interacting with the service. Furthermore, we see significant potential in the explicit consideration of application-specific intents within future IARS.

The paper is organized as follows. Next, in Section~\ref{sec:terminology-and-methodology}, we review different notions of intent-awareness as found in the literature, and we delineate our work from related streams of research. Furthermore, we describe our methodology of identifying relevant papers and discuss the relationship of our work to other surveys. In Section~\ref{sec:categorization}, we then categorize and discuss different technical approaches to build intent-aware recommender systems. Section~\ref{sec:evaluation} subsequently provides a landscape of current research in terms of the used types of data, evaluation approaches, and application domains. The paper finishes with a discussion of research gaps and future research directions.

\section{Terminology and Research Methodology}
\label{sec:terminology-and-methodology}
In this section, we first elaborate a definition of intent-aware recommender systems based on how they are actually reported in the literature. Then, we explain our research methodology and discuss previous works.

\subsection{Terminology and Definitions}
According to the Cambridge Dictionary\footnote{\url{https://dictionary.cambridge.org/}}, the noun `intent' refers to ``\emph{the fact that you want and plan to do something}'' or ``\emph{the intention to do something}''. A similar action-oriented definition is provided by Wikipedia\footnote{\url{https://en.wikipedia.org/wiki/Intent_(disambiguation)}}, where intent is described as ``\emph{an agent's specific purpose in performing an action or series of actions.}'' Common synonyms include the terms `purpose', `plan', `goal', or `aim'. We note that the word `intention' is often used synonymous with `intent' as well, even though the two terms are not entirely interchangeable, with the term `intent' suggesting a ``\emph{clear[er] reasoning or great deliberateness.''}\footnote{\url{https://www.britannica.com/dictionary/eb/qa/intent-and-intention}}
Intuitively, an intent-aware RS should therefore take into account \emph{why} a user is interacting with a system or which goal she or he wants to achieve, thus, complementing the traditional approach of predicting which items might be relevant for the specific user.

\paragraph{Notions of Intent in the Literature.} In the surveyed literature researchers rarely specify explicitly what they mean by the terms `intent' or `intent-awareness' or refer to humanities and social sciences literature, e.g.,~\cite{wang_intention2basket_2020,wang_learning_2022}. More frequently, they provide intuitive examples to illustrate their notion of an intent-aware system. Different interpretations of intent-awareness can be found, and these interpretations are typically tied to the technical approaches that are presented in the respective papers implementing intent-awareness in a system; we will discuss these technical approaches in more detail in Section~\ref{sec:categorization}. Following the spirit of early diversity-oriented IARS approaches, the authors of \cite{kaya_comparison_2019} for example state that ``\emph{[IARS] ensure that the set of recommendations contains items that cover each of the user’s interests [...]}.''
More recent works like~\cite{wang_disentangled_2020} relate intents with observed actions and provide examples, in this case for shopping intents in the e-commerce domain, e.g., ``\emph{passing time}'' or ``\emph{shopping for others}''. In several other works, e.g.,~\cite{chen_improving_2020,chen_air_2019}, the term `intent' is mostly considered to be equivalent with (observed) \emph{interest}, e.g., in specific topics or item categories.
Finally, in some works like~\cite{jannach_session-based_2017}, the term (short-term) `intent' is merely used to describe a sequential recommendation approach that puts more emphasis on the last observed user interactions. Thus, intent is sometimes solely equated with observed user actions, without additional consideration of users' underlying motivations for these actions.

\paragraph{Explicit Intent, Latent Intent, and Implicit Intent Models}
Given the examples from the previous paragraph, we can roughly categorize existing IARS approaches into three classes. First, a comparably small number of the surveyed works are based on a small set of pre-defined and application-specific \emph{explicit} user intents, e.g., ``\emph{discover new music to listen right now}'' or ``\emph{find background music for listen to}''~\cite{mehrotra_jointly_2019}. In such an approach, one can for example apply different recommendation strategies depending on the currently assumed intent. Second, some other works provide examples of such explicit intents for illustration, e.g., ``\emph{shopping for others}''~\cite{wang_disentangled_2020}, but are actually based on the concept of \emph{`latent'} intents. In fact, much of the surveyed literature is based on the idea of incorporating (additional) latent variables in the models to be able to account for a typically small set of possible user intents. Third, there are a number of approaches, as mentioned above, in which intent is considered to be roughly equivalent to observed past \emph{interests} in categories, certain item features, or even individual items. In such approaches, additional variables are commonly introduced, which are however not assumed to directly correspond to specific underlying intents in the sense of goals a user wants to achieve. Instead, such variables may, for instance, model the user's time-varying interest in certain item categories or other assumed relationships between individual items. We denote such approaches \emph{implicit} intent models.

A potential advantage of approaches that are based on explicit intents is that the suitability of a given set of recommendations for a particular intent can be more easily assessed. Furthermore, explicit intent-related information may be used to explain recommendations to users. When using a latent-intent modeling approach, this is not easily possible and it may even remain unclear if the latent intents have any strong correspondence with the intents that users may actually have in a given application setting. On the other hand, relying on latent intents enables the design of application-independent approaches and avoids incorporating application-specific knowledge into the models.

We recall here that the general problem for all three categories is that the \emph{true} (current) intent of a given user is usually unknown, and is ``reconstructed'' or predicted from observed user actions. Notably, this also holds for approaches that are based on pre-defined explicit intents, which remain unknown unless users are explicitly asked about them.

\paragraph{Definition}
Our discussion shows that various notions of intent-awareness can be found in the literature. Some of these approaches explicitly refer to and model the users' underlying intentions and goals, whereas others consider the users' intents in an implicit way, e.g., by relating the users' last observed (category) interests to the underlying user motivations. In order to provide an \emph{inclusive} definition of IARS, we characterize such systems as follows:

\begin{quotation}
 ``\emph{An IARS is a recommender system that is designed to capture the users' underlying current motivations and goals in order to support them. }''
\end{quotation}

Our teleological definition is purposely rather broad, yet it expresses that the design of any IARS should be driven by the ambition to support the users' \emph{current} motivations. We emphasize that we use the term `intent' to exclusively refer to the user's immediate or short-term motivations. Users may certainly have diverse preferences and needs over time. However, the core goal of an IARS is to identify the current need or intent when a user interacts with the recommendation service.

Given the breadth of different notions of intent-awareness mentioned above, a multitude of technical ways exist to \emph{implement} intent-awareness. On one end of the spectrum, there are approaches based on an explicit list of pre-defined application-specific intents. In contrast, there are also works in which a sequential recommender system that puts more emphasis on the most recently observed interactions is seen as a mechanism to implement an (implicitly) intent-aware system, e.g.~\cite{jannach_session-based_2017}.

One might argue that \emph{any} sequential next-item prediction model may be thought of as being intent-aware. We however recall that our definition is based on the design goals of a model. Let us consider Amazon's ``\emph{Customers who bought \dots also bought}'' approach as an example for a basic next-item recommendation technique. The recommendations returned for a given item by this purely statistical approach may either be alternatives (similar movies on a video-on-demand service) or complementary items (accessories for a smartphone on an e-commerce shop). In our interpretation, such an approach would therefore \emph{not} be considered intent-aware. The reason is that there is no element in the underlying design of the approach that aims at capturing the user's current motivations or intent, i.e., whether the user looks for alternatives or complementary items.

\paragraph{Relation to Other Areas}
Besides the described relation to sequential and session-based approaches, also time-aware recommendation techniques~\cite{DBLP:journals/umuai/CamposDC14}---and systems that consider user preference drifts over time---may be used \emph{to implement} intent-awareness in a recommender systems. Furthermore, one may use \emph{contextual} information~\cite{Adomavicius2015} to predict the current user's intent and adapt the recommendations accordingly.\footnote{We consider context-aware recommender systems to be different from intent-aware systems, see also~\cite{rafique_developing_2023}. Context information usually describes the external situation of a user, e.g., their location or the current time, whereas user intent is referring to the `inner' motivations of the user. Contextual factors may certainly influence the user's intent.}

Finally, some early intent-aware approaches follow a personalized diversification approach, e.g., in the form of calibration~\cite{kaya_comparison_2019}, to ensure that the provided recommendations cover a larger number of \emph{possible} current user intents. We iterate here that our definition of IARS is not so much focused on the technical way of implementing intent-awareness. Instead, it targets at the underlying ambition when designing a recommender system. Following this argument, an RS that implements a calibration approach \emph{may} be designed to support intent-awareness, but it may also be merely designed to increase the diversity or reduce the popularity of recommendations on an aggregate level, independent of an individual user's current intent.

We note that the concept of user intent can also be found in the context of conversational recommender systems (CRS)~\cite{jannach2021survey}. There, however, intent (and the problem of \emph{intent detection}) refers to what we may call ``micro intents'' during a conversation. These micro-intents rather relate to what the user wants to express (e.g., ask for an explanation, reject or accept a recommendation) in an ongoing dialogue, see also~\cite{kang_understanding_2017}, than to their underlying motivation to interact with the CRS in the first place. The relation of such works to our present work on IARS is therefore limited.

\subsection{Research Methodology}
\paragraph{Identification of relevant papers}
We adopted a semi-systematic approach to identify relevant papers. First, we queried services like Google Scholar and digital libraries for scientific papers that mention the terms `intent' or `intention' together with the term `recommend' in the title or abstracts. We then screened these papers for relevant ones and followed the references mentioned in these papers to identify additional papers. We only retained papers that had undergone a peer-review process, i.e., we excluded preprints from our analysis. Furthermore, a paper was considered relevant if its contribution was motivated by the idea of intent-awareness in the sense of our definition from above even though it did not explicitly contain our search terms in the title or abstract. As such, a small number of papers was considered that emphasized on identifying and addressing the user's current aims and interests.
We also identified a number of works which mention the term `intent' in the title or abstract, but not in a context that is relevant for our survey. For example, some works aim at predicting if a consumer will make a purchase in an ongoing session or not, e.g.,~\cite{requena_shopper_2020}. In such works, the goal is not to make recommendations that match a certain intent, as done in IARS, but to assess the \emph{general propensity} of a consumer to make a purchase, see also~\cite{koniew_classification_2020}.

Overall, we ended up with 85 papers that we considered relevant and which we categorize later in Section~\ref{sec:categorization}.
A possible limitation of our work is that our survey is not based on the strict procedures of a systematic review in the sense of Kitchenham et al.~\cite{Kitchenham2009,Kitchenham2013ASystematic}. Nonetheless, by using relevant queries and by applying a snowballing procedure to find additional papers, we are confident that our survey comprises a representative collection of papers reflecting the main areas of IARS research and capturing the conceptual breadth of the field.
Our literature search also surfaced surveys that did not propose a technical contribution or reported an experimental study. We discuss these papers separately next.

\paragraph{Relation to other surveys}
Surveys on related the areas, as discussed above, such as time-aware, contextual, and sequential recommender systems, can be found in~\cite{DBLP:journals/umuai/CamposDC14,bogina2022considering,Adomavicius2015,Kaminskas:2016:DSN:3028254.2926720,QuadranaetalCSUR2018,wang2021survey}. Early intent-aware recommendation systems were inspired by existing approaches to search result diversification in the field of Information Retrieval, see~\cite{Santos2015Search} for a related survey. Indeed, considering the users' intents, e.g., in terms of their `informational needs' has a long history in Information Retrieval, see~\cite{Broder2002Taxonomy} for an early taxonomy. A review of intent-awareness in the particular context of multimedia information retrieval can be found in~\cite{kofler_user_2017}. Like in our survey, Kofler at al.~\cite{kofler_user_2017} found that various notions of intent-awareness exist in the literature. Our definition of intent-awareness is similar to theirs by focusing on the `why' dimension, i.e., the reason and goal behind a search, and not the `what' dimension of information needs. In the area of music information retrieval, Schedl et al.~\cite{schedl_current_2018} reviewed current challenges and visions in music recommendation research. While their review work is not primarily focused on the topic of user intents, they identify an improved understanding of listening intents and the user's purpose as a grand challenge in this area. In that context, they also mention that creators may have specific intents in mind when designing a playlist (e.g., for relaxing).

Probably most similar to our survey is the work by Rafique et al.~\cite{rafique_developing_2023}, which also focuses on intent-aware recommender systems. They, however, study the current and future role of IARS from the lens of smart cities. A particular emphasis is put on the role of recent technological developments in this context, e.g., the increasing usage of wearable and intelligent devices, the Internet of Things or Edge Computing approaches. As a result, new opportunities arise to build future intent-aware systems based, for instance, on real-time activity, emotion tracking or the explicit elicitation of user intents based on new sensors and devices. In their survey, the authors thus particularly focus on case studies of IARS in smart cities and on key requirements in this context.

\section{Technical Approaches for IARS}
\label{sec:categorization}
\begin{figure}[h!t]
  \centering
  \includegraphics[width=.7\textwidth]{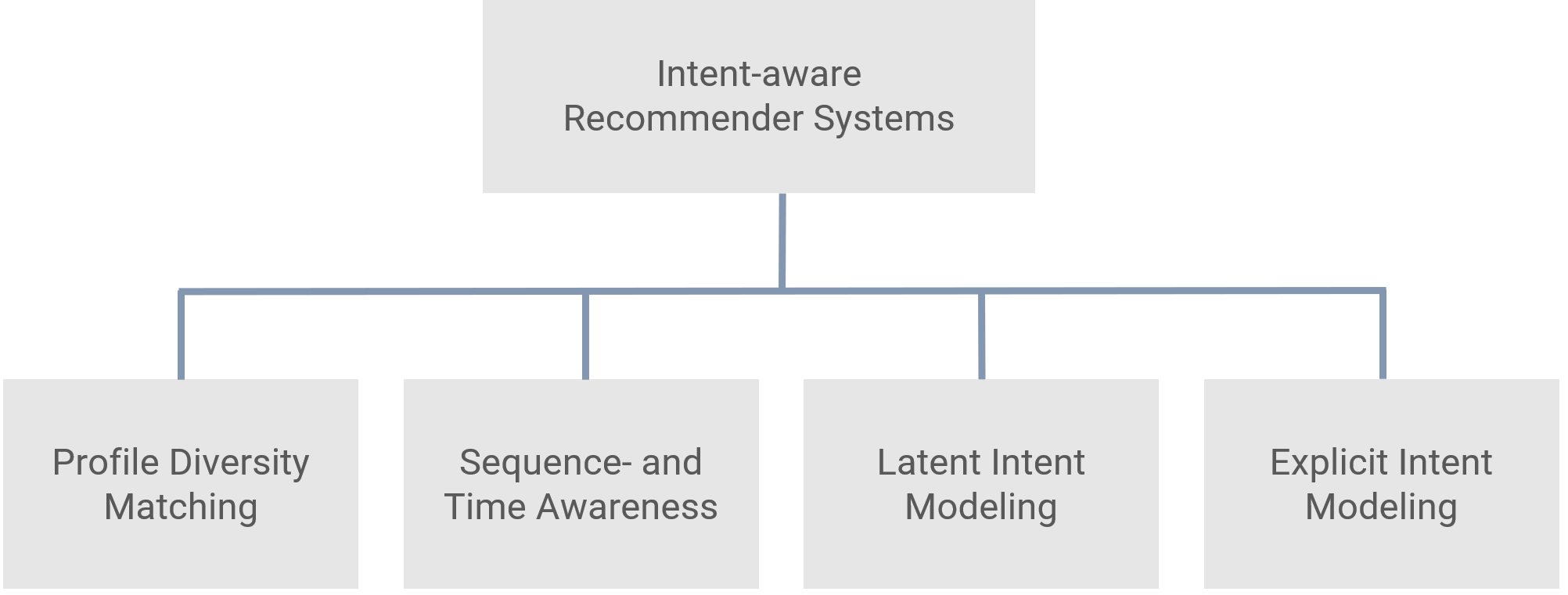}
  \caption{Overview of Technical Approaches}\label{fig:taxonomy}
\end{figure}

In this section, we categorize the considered studies on intent-aware recommender systems according to their distinguishing algorithmic properties. Early \emph{Profile Diversity Matching} approaches have a particular optimization goal and rely on item-metadata. \emph{Sequence- and Time-Aware models} are based on the availability of temporal information about past user interactions. Finally, \emph{Latent Intent} and \emph{Explicit Intent Modeling} approaches assume the existence of a small set of common intents in a given application use case. Our resulting categorization is illustrated in Figure~\ref{fig:taxonomy}.

We note that some papers may fall into more than one category, e.g., a latent-intent modeling approach with the goal of diversification. In such cases, we categorized these works according to what we identified as their main conceptual contribution.

\subsection{Profile Diversity Matching}
\label{subsec:profile-diversity-matching}
The general goal of the approaches in this category is to generate diversified recommendations that cover the spectrum of past user preferences as good as possible. Such approaches are often designed under the assumption that the current user intent cannot be reliably estimated. Thus, if the provided recommendations cover multiple past intents (e.g., multiple categories of items), the chance increases that at least some of the provided suggestions are a match for the current user intent. Since the current user intent is not explicitly taken into account, such approaches, according to our above definition, fall into the category that considers users' short-term needs in an implicit way.

Early works in this area were inspired by previous efforts towards \emph{search result diversification} in the field of information retrieval~\cite{Agrawal2009Diversifying,Santos2012Exploting}. Besides avoiding redundant search results, one specific goal of search result diversification is to account for possible ambiguities of the provided search terms. The provided search results should thus match more than one of the possible interpretations (and underlying search intents) of a given query.

Vargas et al.~\cite{vargas_intent-oriented_2011,Vargas2012Explicit,Vargas2013ExploitingOAIR} were probably one of the first to transfer this approach to the field of recommender systems.  Specifically, in their ``intent-oriented'' diversification approach~\cite{vargas_intent-oriented_2011}, they establish an analogy between a user profile in a recommender system and a search query in a retrieval system. The assumption is that recommender system users often have interests in various things over time, but not all of them are relevant at a given point in time. Technically, their approach is based on the deriving an \emph{aspect-oriented} user profile based on explicit or latent item features. Then, a greedy re-ranking strategy is used to diversify a given accuracy-optimized recommendation list in a way that the various past user interests in the different aspects of the items are taken into account. As a result, this re-ranking procedure aims to avoid that a set of recommendations only covers one particular type of items, e.g., shopping items of only one category.

Several later works built on these ideas. In~\cite{wasilewski_intent-aware_2016}, for example, the authors further explore the use of explicit vs.~implicit aspects for intent-oriented diversification and propose a new model that maintains interpretability and leads to a better accuracy-diversity tradeoff. Tomeo et al.~\cite{Tomeo2015Exploiting} consider the case of multiple relevant item attributes and put forward a method that can both lead to increased diversity at the level of individual user as well as at the aggregate level.

Later Kaya and Bridge~\cite{Kaya2018Accurate,kaya_comparison_2019} propose another re-ranking based intent-aware diversification approach, which is however not based on item features, but on user \emph{subprofiles}. In their approach, subprofiles are automatically extracted sets of items from a given user profile, where each subprofile represents a distinct user taste. Differently from previous aspect-oriented works, the set of aspects is thus not the same across all users but user-individual.

Generally, a central aspect of the discussed approaches is that they aim to diversify recommendations at the level of individual users rather than to maximize diversity at an aggregate level across users, which is the focus of various other research works on beyond-accuracy quality factors~\cite{Kaminskas:2016:DSN:3028254.2926720}. As such, these approaches are related to recent technical approaches for \emph{calibrated} recommendations~\cite{DBLP:conf/icdm/OhPYSP11,JugovacJannachLerche2017eswa,steck2018calibrated}. In calibration approaches, which are also often re-ranking based, the aim is to generate recommendations that well reflect the distribution of different item aspects (e.g., genres of a movie) in a given user profile. Intent-aware diversification and calibration approaches are therefore strongly related, with the main difference that diversification is not necessarily an \emph{explicit} goal of calibration approaches~\cite{Kaya2019Subprofile}.

Two recent and technically alternative approaches that try to account for multiple interests in user profiles are discussed in~\cite{lian_multi-interest-aware_2021} and~\cite{liu_octopus_2020}. The Octopus system presented in~\cite{liu_octopus_2020} focuses on the \emph{candidate generation} phase of modern large-scale recommender systems~\cite{Covington:2016:DNN:2959100.2959190}. Differently from common deep learning recommendation models, Octopus creates multiple vectors for representing a user to model the user's interests in a comprehensive manner. The system proposed in~\cite{lian_multi-interest-aware_2021} adopts similar ideas and also relies on multi-channel networks, but focuses on sequential recommendation problems. A specific aspect of the works proposed in~\cite{lian_multi-interest-aware_2021} and~\cite{liu_octopus_2020} is that they were designed for large-scale applications. The work in~\cite{lian_multi-interest-aware_2021} was furthermore field-tested in an online experiment, where it showed to lead to a marked increase in user clicks and revenue.

In terms of real-world applications, little is published in the literature on how large-scale platforms like YouTube technically diversify their recommendations to cater for time-varying user intents. Various real-world online services, e.g., Netflix or Amazon, however apparently address this problem by providing multiple recommendation lists to users at a time~\cite{Gomez-Uribe:2015:NRS:2869770.2843948}. At least some of these lists are clearly connected to a particular assumed user intent, e.g., ``continue watching'' or ``discover our new arrivals''. Even though such multi-list approaches are common in practice, only a few works on the topic can be found in the academic literature, e.g.,~\cite{jannachmultirow2021,Starke2021Serving,Rahdari2022TheMagic}.

\subsection{Sequence- and Time-Aware Approaches}
\label{subsec:sequence-and-time}
Approaches in this category leverage temporal or sequential information in user behavior logs to provide recommendations that assumably match the current intent of users. Commonly, the most recent observed user interactions are considered to be indicators or representatives for the underlying user needs or goals. In a broad interpretation of this definition, all session-based, all sequential and many time-aware recommender systems---see~\cite{wang2021survey,QuadranaetalCSUR2018,DBLP:journals/umuai/CamposDC14,bogina2022considering} for related surveys---may thus be seen as intent-aware approaches. In the following, we will review a set of selected works, which in one or the other way mention intent-awareness as a motivating element for their model design.

\paragraph{Session-based Approaches} In the context of \emph{session-based recommendation}, where item suggestions are not based on long-term preference profiles but only the interactions that observed in an ongoing session, the models NARM~\cite{li_neural_2017} and STAMP~\cite{liu_stamp_2018} are sometimes referred to as being intent-aware. Both models use an attention mechanism to focus the recommendations on the assumed main user intent in a given session. In the case of STAMP, the potential user interest drift during a session is taken into account and a particular focus is put on the most recent interactions in a session. In this context, the authors differentiate between long-term and short-term signals, even though only the interactions of individual sessions are considered.\footnote{In other works, e.g.,~\cite{jannach_session-based_2017}, the term long-term interests is used to denote interaction data from previous sessions of the same users.} Generally, the need for identifying the current main intent may arise from the problem that a user might have multiple intents during one single session; plus, there can be noise, e.g., random interactions, in the logged user behavior.
Another related approach in session-based recommendation can be found in~\cite{Wang2019ACollaborative}, where a sequence modeling approach is combined with a neighborhood-based model, which considers similar sessions by other users that are assuming the same intent. Neighboring sessions are also considered in the `intent-guided' approach in~\cite{pan_intent-guided_2020}. This follows a two-stage approach, where in the first phase, like STAMP, the focus lies on the most recent interaction to derive a representation of the assumed intent. This information is then used to select suitable neighbor sessions to compute final predictions. A quite different approach to intent-modeling in session-based recommendation is proposed in~\cite{guo_learning_2022}. Specifically, in their graph-based approach, Guo et al.~aim at extracting and leveraging user intent signals from `groups of consecutive items' at different levels of granularity. A graph-based approach is also proposed by Li et al.~\cite{li_spatiotemporal-aware_2022}, who particularly focus on the integration of spatiotemporal information for improved session-based recommendation.\footnote{In some ways the approach in~\cite{li_spatiotemporal-aware_2022} has elements of a latent-intent model technique as they rely on some form of intent embeddings. We iterate that the categorization of existing approaches in this section is not always entirely exclusive.} Finally, Jin et al.~\cite{jin_dual_2023} focus on the \emph{new item} problem in session-based recommendation, i.e., on the recommendation of items for which no past interaction data is available. The basis for their graph-based approach lies on the existence of a detailed taxonomy of categories and other side information on items.

\paragraph{Session-Aware Approaches} In \emph{session-aware}~\cite{QuadranaetalCSUR2018} recommendation systems, the logged user interactions are also organized in sessions. However, differently from pure session-based approaches, information about previous sessions of the current user are available, allowing for \emph{personalized session-based recommendations}~\cite{quadrana17personalizing} based on long-term user preferences. In an early work in this area, Ludewig et al.~\cite{jannach_session-based_2017} investigate the relative importance of long-term preferences and `short-term intents' in the fashion e-commerce domain. Their results show that both long-term and short-term preferences signals can be relevant. Focusing strongly on the last few observed interactions is however essential in this domain. Additional analyses in this work also revealed that \emph{reminding} consumers to continue their previous shopping session, as another typical intent of online shoppers, can be highly effective. Later, Huang et al.~\cite{huang_attention-based_2018} propose the ASLM model to combine information from previous sessions with a user's current session in a deep learning approach. The model, which relies on Recurrent Neural Networks (RNN) and the attention mechanism, combines a long-term preference layer with a short-term intent layer to obtain improved performance. In a related study, Bernardis et al.~\cite{bernardis_data_2022} compare different session-aware recommendation models based on a real-world dataset from a video-on-demand online service. Their results indicate that considering long-term preferences in the examined RNN-based model can be beneficial, but the additional gains may be quite limited in this particular application setting. A technically quite different approach to combine historical user sessions and the current user intent is proposed by Lui et al.~\cite{liu_intent_2020-1}, taking into account practical constraints in online settings. They propose a generic learning framework that decouples the process of \emph{online} learning of the current intent (in the ongoing session) and the learning process for the general user preferences (from past sessions). Moreover, to be able to quickly learn the current intent during a browsing session, a meta-learning approach is adopted instead of online learning techniques that have been previously used to deal with the streaming feedback on large online platforms. An alternative approach based on memory networks to combine long-term and short-term preferences is proposed in~\cite{zhu_sequential_2020}. In this work, the users' short-term intents are mostly equated with their interest in specific item categories, leading to a hierarchical user model. Later on, this work was further developed by the authors in~\cite{zhu_learning_2022}. The improvement of the new model is that category information is no longer required, but multiple intent levels can be modeled instead.

\paragraph{Sequential Approaches} A number of intent-aware approaches can also be found in the area of \emph{sequential} recommendation settings. Like in the previously discussed session-based scenarios, sequential recommendation approaches are based on time-ordered user interaction logs and the general goal is to predict the user's immediate next action. However, these approaches do \emph{not} assume that the data is organized in sessions. The differentiation of long-term preferences and short-term intents is central to several works also in sequential recommendation settings. Zhang et al.~\cite{zhang_next_2019}, for instance, propose a corresponding architecture that comprises two components to capture a user's \emph{transient interests} based on their observed interaction trajectories: a self-attention model for short-term intents and a latent factor model for long-term preferences. Long-term and short-term aspects are also combined by Zhang et al.~\cite{zhang_next_2018} and by Liu et al.~\cite{liu_tpgn_2021}, where the latter develop a novel approach to consider time intervals between observed events. Other approaches involve sophisticated temporal reasoning approaches as well as model the evolution of user behavior over time. These models may additionally also rely on item meta-data to factor in semantic relationships between items~\cite{wang_toward_2021} or time-varying popularity of items~\cite{jiang_dynamic_2021-1}. The use of item meta-data, in particular in the form of \emph{category} information, is central to a number of other works~\cite{guo_intention_2020,chen_sequential_2022,chen_air_2019,xu_modeling_2021,cai_category-aware_2021-1}. In many of these works, the time-varying interest in different item categories is used as a proxy for user intent, and the goal is thus to model item category trajectories in the interaction data. Technically, graph-based approaches and attention are commonly used. Moreover, different types of side information are taken into account, including social information~\cite{chen_sequential_2022}, purchase cycles~\cite{guo_intention_2020} or information extracted from tags or user reviews~\cite{li_intention-aware_2022}.
Some works also consider multiple possible category-related intents in parallel~\cite{xu_modeling_2021} and are able to process multiple types of user actions (e.g., clicks, purchases etc.)\cite{chen_air_2019,xu_modeling_2021}. Finally, a few proposals exist that consider \emph{domain-specific} aspects. Zhu et al.~\cite{Zhu2023ikgn}, for example, focus on Point-of-Interest recommendation problems and their model incorporates a ``locational intent'' signals from spatio-temporal information and location categories. Ding et al.~\cite{ding_modeling_2022}, on the other hand, concentrate on the fashion domain. They consider shopping-specific intents (e.g., ``match'' or ``substitute'') and furthermore incorporate automatically extracted features of fashion items, which are used to model item sequences on a level not tied to individual items. Lastly, two intent-aware approaches for predicting the next app usage on mobile phones are presented in~\cite{lee_making_2014} and~\cite{changmai_-device_2019}. In~\cite{lee_making_2014}, a system based on (sequential) rules is proposed that predicts the next user action, i.e., app invocation, based on the user's predicted intent and context. A related approach is presented in~\cite{changmai_-device_2019}, which relies both on spatio-temporal and sequential information to predict the next user intent.

\subsection{Latent Intent Modeling}
\label{subsec:latent-intent-modeling}
A common assumption of papers in this category is that multiple underlying \emph{latent} intents can be behind an observed user-item interaction. Correspondingly, these papers propose neural models that consider these intents through additional latent variables or additional layers in the network. In most cases, assuming that a limited number (e.g., 4 to 12) of such latent intents exists, leads to the best results. Differently from the works that will be discussed later in Section~\ref{subsec:explicit-intent-modeling}, which are based on knowledge about application-specific intents, the semantics of the intents are not known in latent intent modeling approaches. We grouped the identified works into different categories, as described below.

\paragraph{Disentanglement Approaches using Factorized Representations}
Several of the identified works aim at modeling the underlying user intents through disentangled representations~\cite{Bengio2013Representation}. In such factorized representations, ``\emph{a change in a single unit of the representation corresponds to a change in single factor of variation of the data while being invariant to others}''~\cite{Dupont2018Learning}. Technically, the goal in these approaches commonly is to learn a \emph{chunked} user or item representation~\cite{zhao_multi-view_2022}, where each part of the representation corresponds to an intent.\footnote{Technically, we note that factorized representation approaches are related both to capsule networks and multi-head attention networks, as discussed in~\cite{wang_disentangled_2020}. Furthermore, disentanglement approaches have been proposed for recommender systems for other purposes than intent-awareness, e.g., in~\cite{zhang_content-collaborative_2020}.}

Ma et al.~\cite{ma_learning_2019} were among the first to propose to learn disentangled representations for recommender systems. Their approach named MacridVAE combines macro-level and micro-level disentanglement. At the macro level, the user's interest in a limited set of \emph{k} high-level concepts associated with user intents is modelled, and each component of the user representation captures the user preference regarding the \emph{k$^{\text{th}}$} concept. A micro-level regularizer then ensures that each element of the user representation independently reflects a more low-level item factor, e.g., a certain item attribute. In the same year, Ma et al.~also propose a general disentanglement method called DisenGCN for graph representations~\cite{pmlr-v97-ma19a}. Wang et al.~\cite{wang_disentangled_2020} then apply this principle of disentanglement for graph representations for collaborative filtering problems, leading to the DGCF method. Specifically, their proposed method includes the construction and iterative refinement of multiple intent-aware graphs to propagate information to the intent-aware chunks in a fine-granular way. Also, a dedicated component of the proposed frameworks guides the learning in a way that the factor-aware representations are independent. A similar method was proposed by Wang et al.~\cite{yu_multi-intent_2022}, who propose to use a transformer network and model the correlation between intents. Some additional works explore disentangled representations for specific problem settings such as bundle recommendation~\cite{zhao_multi-view_2022} or for the particular use case of news recommendation~\cite{hu_graph_2020}. The consideration of further types of side information in intent-aware disentanglement approaches was proposed in different recent works. Wu et al.~\cite{wu_intent-aware_2022}, for example, leverage tag information in a self-supervised learning approach. Wang et al.~\cite{wang_causal_2022}, on the other hand, devise a causal method for Heterogeneous Information Networks (HIN) that may contain information about item attributes or the social relationships of the users. Finally, a disentanglement approach in which item representations are sliced in different intent-related chunks was proposed by Li et al.~\cite{li_enhancing_2022} for the class of session-based recommendation problems~\cite{Jannach2021SessionRSHB}.

Besides the discussed approaches that rely on factorized representations, a number of alternative latent-intent modeling approaches were put forward in the literature, both for traditional top-N recommendation and for sequential recommendation problems. We discuss these in the next paragraphs.

\paragraph{Alternative Latent-Intent Modeling Approaches for Top-N Recommendation}
Rather than slicing the user item representations into different chunks, typical works in this category propose architectures that learn multiple embeddings, one for each latent intent, or introduce additional layers to learn intent-enhanced embeddings.\footnote{Some of these works also use the term `disentanglement', but the technical approaches are different from factorized representations discussed previously.}

The DisenHAN model by Wang et al.~\cite{wang_disenhan_2020} features a particular disentangled graph attention network that leverages different aspects in a Heterogenous Information Network (HIN). The output of their layered architecture are a number of intent-enhanced embedding vectors, which represent different aspects of a user and items. A related approach to intent-aware recommendation based on knowledge graphs was proposed by~\cite{wang_learning_2021}. The work is conceptually similar to the DGCF approach discussed above~\cite{wang_disentangled_2020}, but was designed to also consider fine-grained item information in the intent-modeling process.  Knowledge graphs and intent-awareness were also in the focus of Zhang et al.~\cite{zhang_fire_2022,zhang_knowledge_2022}. In their models, the user and item representations are disentangled to different spaces, resulting in a number of intent embeddings, which are then used to obtain enhanced embeddings that are finally used for recommendation. Similar ideas were later adopted also by Li et al.~\cite{li_entity-driven_2023}, who create intent graphs for each user before training based on connected entities in the knowledge graph. A KG-based method inspired by \emph{topic modeling} was proposed by Li et al.~\cite{li_topic-aware_2023-1}, which aims to overcome potential problems when extracting intent information from graph relations, as done, e.g., in~\cite{wang_learning_2021}. Finally, Lin et al.~\cite{Lin2023Attention} propose an intent-aware approach that re-ranks and refines a recommendation list by learning latent user intent representations from text reviews. Their work is unique in that it uses user-generated content in the form of reviews as auxiliary information.

A number of other works address specific problem settings. Li et al.~\cite{li_package_2021}, for example, focus on the \emph{package recommendation} problem and propose an approach to create multiple disentangled user embeddings. Wang et al.~\cite{wang_learning_2022} address the \emph{complementary item recommendation} problem and propose an architecture that creates an \emph{intent embedding} based on the user embedding and category information. This intent embedding is then combined with a module that creates aspect-level complementarity embeddings to create the final recommendations. Wei et al.~\cite{wei_hierarchical_2022} concentrate on \emph{multimedia recommendation}, and they propose a hierarchical approach that infers coarse-grained and fine-grained intent levels from observations of co-interacted items. Finally, Qian et al.~\cite{qian_intent_2022} focus on the problem of \emph{popularity bias} in recommendations. Specifically, they aim to disentangle the underlying intent of the user into conformity (to like popular items) and genuine interests, and their approach is based on creating the correspondingly disentangled representations for users and items.

\paragraph{Alternative Latent-Intent Modeling Approaches for Sequential Recommendation}
A variety of proposals have been put forward for latent-intent modeling for sequential recommendation problems in the past few years. Like the previously discussed approaches, these proposals are commonly based on projecting objects into a limited number of intent spaces, on modeling sequential patterns of intents or categories, or on disentanglement techniques.

In one of the earlier works in this area, Tanjim et al.~\cite{tanjim_attentive_2020} propose an intent model that uses self-attention and temporal convolutional networks to identify item similarities from interaction sequences. Given the sparsity of the interaction data, they furthermore propose to consider user actions on the category level, resembling ideas of modeling category transitions discussed earlier. The approach by Di~\cite{di_multi-intent_2022} is similarly motivated by sparsity considerations, and the goal of the proposed architecture is to transform the given interaction sequence from the item space to the intent space. Technically, the work by Di involves a particular contrastive learning (CL) learning scheme. CL is also the core of related approaches by Li et al.~\cite{li_multi-intention_2023} and Chen et al.~\cite{chen_intent_2022}. In~\cite{li_multi-intention_2023}, sparsity is addressed through data augmentation, and a sequence encoder is designed which projects the historical items of a user into a small set of latent spaces, integrates local and global intention representations, and selects the current main intention.
The approach by Chen et al.~\cite{chen_intent_2022} learns the intent distributions from user behaviour sequences via k-means clustering and proposes a new contrastive self-supervised learning objective. The assumption of the existence of only a very limited set of interest categories (concepts) is challenged by Tan et al.~\cite{tan_sparse-interest_2021-1}, who propose a sparse-interest network architecture designed to adaptively activate a subset of concepts from a larger pool of existing concepts for a user.

Various alternative technical approaches to intent-aware sequential recommendation are proposed in~\cite{bhattacharya_intent-aware_2017,cen_controllable_2020,sulikowski_fuzzy_2021,wang_target_2022,li_misrec_2023,ma_disentangled_2020,chang_latent_2023}. Cen et al.~\cite{cen_controllable_2020} focus on the candidate generation phase of recommender system. They explore two methods for multi-interest extraction, one based on self-attention, as done in several other works, and one in which the multiple interests of a given user are viewed as \emph{capsules}~\cite{Sabour2017Dynamic}. In this latter approach, the dynamic routing method from CapsNet is used to generate multiple output capsules that correspond to interest embeddings. Wang et al.~\cite{wang_target_2022} later challenge the effectiveness of the \emph{greedy inference} method used, for example, in~\cite{cen_controllable_2020} and propose a target-interest \emph{distillation} approach to dynamically aggregate multi-interest embeddings for a given context. Li et al.~\cite{li_misrec_2023} instead present another work based on the attention mechanism, and they specifically use multi-head attention, where each head corresponds to one of a limited set of attentions to users. Ma et al.~\cite{ma_disentangled_2020}, like ~\cite{cen_controllable_2020}, focus on candidate generation, and they challenge the effectiveness of multi-head attention in models such as SASRec~\cite{Kang2018sasrec} for intent modeling. Instead, they propose a specific intention disentanglement layer appended after single-head SASRec encoder, which involves an intention clustering and weighting step for a given small set of latent categories. In yet another technical approach, Chang et al.~\cite{chang_latent_2023} put forward a probabilistic approach in which user intent is modeled as latent variables, which connect past observed user behavior and future behavior. Oh et al.~\cite{oh_implicit_2022}, on the other hand, propose to model what they call ``implicit session contexts'' in session-aware recommendation settings. An implicit session context is seen to be equivalent to a latent intent, and the model uses a next-context predictor to guide the next-item prediction model.

The works by Bhattacharya et al.~\cite{bhattacharya_intent-aware_2017} and Sulikowski et al.~\cite{sulikowski_fuzzy_2021} stand out in that they use very specific ways of considering the users' past and ongoing interaction behavior for intent-modeling and prediction. Bhattacharya et al.~focus on a particular use case, where textual reports are recommended to users. Their intent modeling approach considers both navigational aspects, e.g., which reports users have interacted with and for how long, as well as content-related aspects. A particularity of their use case is that the authors equate the item to be recommended, i.e., a specific report, with the predicted intent or user goal. This aspect is similar to the work by Sulikowski et al., who aim to predict the intent of users to purchase individual items, which is similar to the classical recommendation problem. However, the purchase probabilities in their work are based on an intent-modeling approach that considers fine-grained user interaction signals such as clicks, dwelling times, mouse moves, and scrolling activities. A further unique aspect of their approach also lies in the use of a fuzzy rule-learning technique.

While most identified works aim to increase recommendation accuracy through intent-awareness, a few works also explicitly address questions of recommendation diversity. The approach by Chen at al.~\cite{cen_controllable_2020} discussed above, for example, incorporates a factor to control accuracy and diversity when integrating items from different intents. Achieving high diversity in terms of categories while preserving accuracy is a central goal in the intent-aware approach by Chen et al.~\cite{chen_improving_2020}. Technically, diversity aspects are considered as part of a multi-element loss function. The balance of diversity and accuracy is controlled by a hyperparameter, as done in many other works. A different technical approach to achieve category-wise diversity is followed by Wang et al.~\cite{wang_modeling_2019}. They propose a multi-channel approach, where each channel is implemented through a recurrent neural network and corresponds to a specific user purpose (intent). These different channels are then used to create diversified recommendations assumed to match the user's multiple intentions in a given session.

Finally, a few works in the literature target the specific problem of next-basket recommendation. In~\cite{wang_intention2basket_2020,wang_intention_2020}, Wang et al.~base their work on psychological theories and split the basket prediction problem into the phases of intent recognition, modeling, and accomplishment. The technical approach includes both computing the probability that a given intention is driving a certain observed user choice and to learn intention transition patterns in the data. The learned intents are finally also considered for the construction of the basket. A quite different technical approach for basket recommendation is proposed by Lio et al.~in \cite{liu_basket_2020}. Specifically, a graph-based and translation-based approach~\cite{Bordes2013Translating} is applied to generate multiple representations for a basket according the a pre-defined number of latent intents. One goal is then to learn the importance of each intent for a given basket. A specific property of the approach is that it is able model also the correlation between intents.

\subsection{Explicit Intent Modeling}
\label{subsec:explicit-intent-modeling}
In this category, we discuss research works that address intent-awareness for particular use cases. These works thus makes assumptions regarding the existence of specific intents in a certain domain.

In the media domain, Kapoor et al.~\cite{kapoor_i_2015} found that users of music services may have time-varying novelty preferences. They engineered a set of features to predict a user's varying appetite for new items over time, and incorporated these predictions into a recommendation algorithm. Overall, with their approach, the authors focus on exactly one particular intent, i.e., to discover new artists or tracks. Volokhin and Agichtein~\cite{volokhin_understanding_2018,volokhin_towards_2018}, in contrast, aimed at understanding more generally during which activities and with which intents users consume music. Through a survey-based research they identified seven main intents (concentration, distraction, filtering background noise, inspiration, mood and emotion control, motivation, relaxation).\footnote{We note that this number of intents matches the typical range of intents in latent-intent modeling discussed above quite well.} Initial experiments furthermore indicate that these intents may be connected with audio characteristics of the recommendable tracks, ultimately leading to better recommendations. A number of music-specific intents was considered also in the study by Mehrotra et al.~\cite{mehrotra_jointly_2019}. Differently from the work in~\cite{volokhin_understanding_2018}, however, the authors asked survey participants about their intent when accessing the homepage of the music service (as opposed to the intent of listening to music in general). The identified intents are therefore partially of a different nature, and include intents such as ``to quickly access my playlist [\dots]'', ``to discover new music to listen to'', or ``to find music to play in the background''. These intents are then used to develop a multi-level model that predicts user satisfaction based both on interaction signals and user intents. Mehrotra et al.'s study was later replicated by Benedict et al.~\cite{benedict_intent-satisfaction_2023} for the video streaming domain. In this replication study, eight intents were identified, which the authors grouped into two categories, `explorative' (e.g., finding something new) and `decisive' (e.g. looking for a specific title). The analyses by the authors show that intent affects user behavior and satisfaction also in this domain.

In the area business-to-consumer of \emph{e-commerce}, He et al.~\cite{he_intent-based_2014} describe a psychology-informed approach to build an intent-aware recommender system that has the specific purpose~\cite{JannachAdomavicius2016} to convert visitors to buyers. A visitor's intent is considered equivalent to their propensity to buy. A Hidden Markov Model for the consumer decision process was designed with these five explicit states: `aware', `interested', `compulsive', `purchase', `abandon'. The predicted state of the user is then used to inform the subsequent selection of the content to be presented. A related approach was later put forward by Shi et al.~\cite{shi_improving_2017}, who identified four possible user intentions in e-commerce shopping from the literature: \emph{research shopping}, \emph{comparative browsing}, \emph{idea searching}, and a \emph{hedonic} intention. Like in~\cite{he_intent-based_2014}, the authors use clickstream data to first predict the current user intent and then select one out of several predefined recommendation algorithms in a switching hybrid approach. Differently from works that model general e-commerce shopping intents, Ding et al.~\cite{ding_modeling_2022} focus on fashion e-commerce and identify three common intents: \emph{match}, \emph{substitute}, and \emph{others}. Technically, a translation-based~\cite{Bordes2013Translating} approach is used to model the interaction between a user, their previously interacted item, and the current intent, with the goal to infer the probability of each intent given these observations. Only two possible intents are considered by Loyola et al.~\cite{loyola_modeling_2017} for a session-based e-commerce recommendation setting: browsing or purchasing. Technically, an encoder-decoder recurrent neural network architecture is proposed, which is augmented with a second decoder that predicts the value of a binary variable that represents the current shopping intent. The existence of two main intents is also the assumption in the work by Sobecki et al.~\cite{sobecki_towards_2014}. In their context-aware approach to recommend shopping places, they differentiate between consumers who are either open for \emph{discovery} (with no particular shopping goal) or where they are interested in \emph{efficiency} (and already have a particular shopping goal). The recommendations in the proposed technical approach are based on rule mining techniques, where the importance of the derived rules is determined by the user's \emph{explicitly specified intent} and the geographical and social context of the user. Explicitly stated intents are also the focus of the study by Yang et al.~\cite{yang_how_2019}. In their work in the context of podcast recommendations, user intents correspond to explicitly stated interest categories, i.e., podcast topics. The goal of their user study was then to find out to what extent a recommendation strategy that is informed by user intents or aspiration can be favorable over an intent-agnostic one.

Finally, a very specific use-case of an intent-aware recommender system is described by Fan et al.~\cite{fan_metapath-guided_2019} and by Yang et al.~\cite{yang_finn_2021}. Both works propose related methods for \emph{intent recommendation} for e-commerce sites. With the term intent recommendation the authors refer to the personalized recommendation of search queries, i.e., the goal is to predict which search query a user should provide to find relevant items, given their currently assumed intent. The search field of the e-commerce site is then automatically pre-filled with such a query.
The approach proposed in~\cite{fan_metapath-guided_2019} relies on a Heterogenous Information Network and on a novel metapath-guided embedding approach. Later on, in~\cite{yang_finn_2021}, a \emph{feedback interactive neural network} is proposed that is \emph{(i)} able to consider both positive and negative feedback, \emph{(ii)} filters out noise and \emph{(iii)} relies on multitask learning to match user search intents with query candidates. A particularity of the discussed approaches is that the predicted query is directly representing the assumed intent.

\section{Research Methodologies}
\label{sec:evaluation}
In this section, we now turn our attention to methodological questions, i.e., how IARS are evaluated in the literature. We analyzed the papers that were identified by our literature search along different dimensions. This allows us to understand the current landscape of evaluation approaches and to identify potential research gaps.

\subsection{Domains, Datasets, and Model Inputs}
\paragraph{Application Domains}
We first categorized the research works in terms of the application domains. As done by previous research~\cite{klimashevskaia2023survey,JannachZankerEtAl2012}, we designed a number of broader categories and classified each paper manually, either based on the datasets that were used for offline evaluation or based on the given application setting in case of human-centric evaluations or field tests.

\begin{figure}
    \centering
    \includegraphics[width=.6\textwidth]{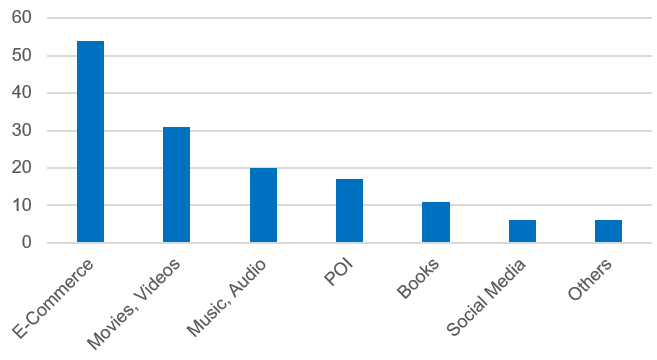}
    \caption{Application Domains of IARS}
    \label{fig:domains}
\end{figure}

Figure~\ref{fig:domains} shows the application focus of current research in IARS. We note that one paper can be assigned to more than one application domain. The analysis shows that \emph{e-commerce} scenarios are the main focus and driver of research in this area. This distinguishes the field of IARS from other surveys, where movie recommendations were traditionally the most common application domain~\cite{klimashevskaia2023survey,JannachZankerEtAl2012}. We attribute the focus on e-commerce domain mainly to the increased interest in session-based and sequential recommendation problems, which are prevalent in e-commerce environments.
Furthermore, the availability of relevant datasets and widely-accepted baseline models may have contributed to the observed focus on e-commerce as an application area.
In the residual \emph{Others} category, we find individual works that for example focus on specific problems such as next-app prediction or ad recommendation.

\begin{figure}
    \centering
    \includegraphics[width=.6\textwidth]{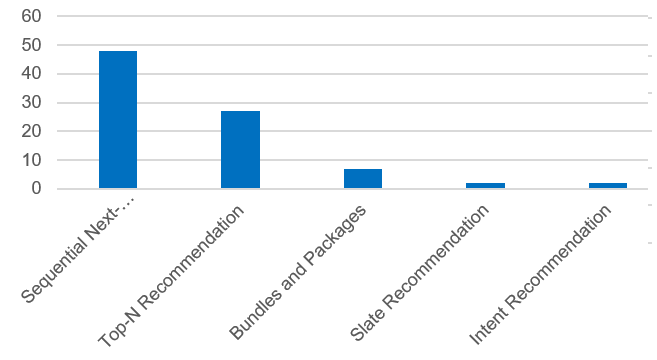}
    \caption{Recommendation Scenarios}
    \label{fig:scenarios}
\end{figure}

Figure~\ref{fig:scenarios} shows the distribution of recommendation scenarios addressed in the surveyed works. The results indeed confirm that sequential recommendation settings (including session-based and session-aware models) are a main driver of research in intent-aware recommender systems. Sequence-agnostic top-N recommendation scenarios are typically in the focus of profile diversity matching approaches and of different (graph-based) intent disentanglement approaches. A few works address package or bundle recommendation problems.\footnote{We note that these are typically also sequential approaches.} Finally, a small number of works focuses on application- specific settings such as slate recommendation~\cite{mehrotra_jointly_2019} or intent (query) recommendation~\cite{yang_finn_2021}.

Interestingly, while many of the surveyed works address problems that are highly relevant in practice---in particular sequential recommendation in e-commerce---research in the area of IARS is still largely driven by academia. Only a handful of papers are authored or co-authored by research groups from industry. Correspondingly, we also did find only a number of papers where IARS models were evaluated in an A/B test in a production environment, as we will discuss later.

\begin{figure}
    \centering
    \includegraphics[width=.6\textwidth]{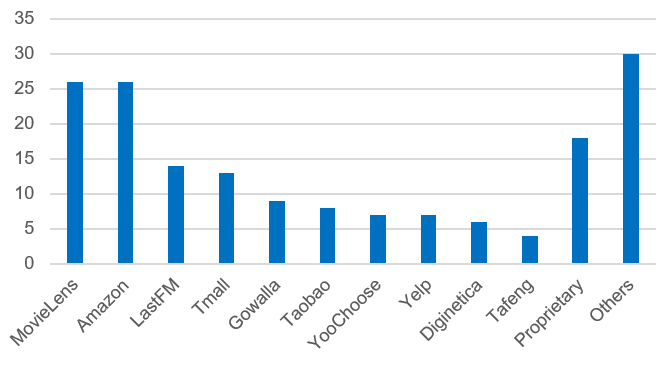}
    \caption{Datasets}
    \label{fig:datasets}
\end{figure}

\paragraph{Datasets}
In Figure~\ref{fig:datasets} we report which datasets are commonly used in the literature. It shows datasets that were used in at least three papers. Most commonly, one of the popular MovieLens datasets\footnote{\url{https://grouplens.org/datasets/movielens/}} or one or several review datasets from Amazon\footnote{\url{https://cseweb.ucsd.edu/~jmcauley/datasets.html\#amazon\_reviews}} are used for (offline) evaluations. As expected, a number of e-commerce datasets that are commonly used to evaluate session-based and sequential recommendations are frequently used by researchers, e.g., from TMall, Taobao or YOOCHOOSE.\footnote{More information about these datasets can for example be found at \url{https://recbole.io/dataset\_list.html}.} The use of proprietary datasets is not uncommon, and frequently, these proprietary datasets contain e-commerce data. Finally, the category \emph{Others} comprises 30 different datasets that only appeared in one or two papers.

\paragraph{Model Inputs}
Except for those few cases mentioned above, where users are expected to explicitly specify their current intent, a given user's intent is assumed to be unknown and must be estimated from the observed user behavior, from contextual factors or from different forms of side information. Considering the surveyed literature comprising 85 papers, the following observations can be made.

In more than 30 papers, i.e., in above one third of the cases, no information beyond the observed interaction signals (e.g., item views, purchases, reviews) and their temporal ordering or timestamps is taken into account. These cases, for instance, include works on session-based and sequential recommendation, where potential user intents are modeled through item co-occurrences in sessions or where the goal is to guess the main intent of a given session, for example through an attention mechanism~\cite{li_neural_2017,liu_stamp_2018,oh_implicit_2022}. Also several approaches that aim at modeling multiple current interests of users, which create multiple intent-based embeddings, or which propose specific network layers for intent modeling, do not leverage any side information, e.g.~\cite{wang_target_2022,zhu_learning_2022}. In terms of the used interaction signals, most works only consider one type of interaction, e.g., item views, even though datasets like YOOCHOOSE or Diginetica exist, which comprise multiple types of user interactions. Interestingly, among all surveyed papers, only a handful consider more than one type of interaction in parallel, e.g.,~\cite{mehrotra_jointly_2019,yang_finn_2021,xu_modeling_2021,loyola_modeling_2017,shi_improving_2017,liu_tpgn_2021}.

Leveraging additional information about users and items beyond interaction signals is however not uncommon in IARS research. About half of the papers incorporate some form of \emph{item meta-data}. Most commonly, the \emph{category} of the items is a central feature, and in several works the observed interest in items of a certain category is used as a proxy for the user's unknown intent. Other types of item meta-data, e.g., the genre of a movie, are commonly used in knowledge graph-based approaches to intent-aware recommendation. Other types of information are only considered in a small set of works,
like \emph{user demographics}~\cite{li_multi-interest_2019,liu_tpgn_2021},
\emph{item content/concepts}~\cite{li_intention-aware_2022,wei_hierarchical_2022,hu_graph_2020},
\emph{user tags and reviews}~\cite{Kaya2018Accurate,wu_intent-aware_2022,guo_intention_2020,volokhin_towards_2018},
\emph{social information}~\cite{li_package_2021,chen_sequential_2022,sobecki_towards_2014,changmai_-device_2019,li_automatically_2022},
\emph{context (location and time)}~\cite{lee_making_2014,li_spatiotemporal-aware_2022,sobecki_towards_2014, zhu_ikgn_2023},
or \emph{item popularity}~\cite{jiang_dynamic_2021-1}.

\subsection{Experimental Approaches}
Commonly, we differentiate between three main ways of evaluation approaches in the literature~\cite{Shani2011}. Offline experiments on historical data, (lab) studies with users, and field tests (A/B) in deployed systems. In addition to these main categories, other types of (non-experimental) research methods can be found from time to time in the general literature on recommender systems, like observational studies or surveys.

\begin{figure}[h!t]
    \centering
    \includegraphics[width=.5\textwidth]{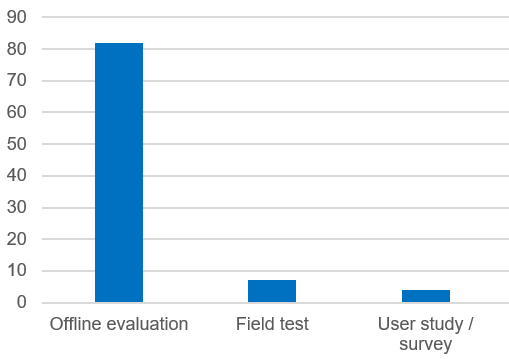}
    \caption{Evaluation approaches}
    \label{fig:evaluation-approaches}
\end{figure}

Figure~\ref{fig:evaluation-approaches} shows the distribution of the main experimental evaluation approaches in the surveyed works. The distribution is well aligned with observations from other surveys~\cite{klimashevskaia2023survey,dejdjoo2023fairness}, where, despite the limitations of this methodology~\cite{jannach2021mcnamara}, the large majority of published research is based on offline experimentation.

The outcomes of field studies with real systems are reported in~\cite{shi_improving_2017,jannach_session-based_2017,li_multi-interest_2019,zhang_efficiently_2023,chang_latent_2023}. The ERIC system presented in~\cite{shi_improving_2017} was evaluated on a Chinese e-commerce site in a two stage approach. In the first phase, online customers were surveyed about their intents, and their behavior was recorded. The collected data was then used to train a model to predict a user's intent from their behavior and to subsequently switch to the most appropriate recommendation strategy. The three months field test showed a significant and consistent improvement in terms of click-through-rate (CTR), purchases, and the average time spent on the recommendations. The comparison in that study was made with a baseline that uses also a hybrid approach but without the intention-based switching strategy. The work by Li et al.~\cite{li_multi-interest_2019} also focuses in the e-commerce domain. Besides offline experiments, the authors tested their capsule network based approach for one week on the TMall homepage. The obtained results showed that their proposed MIND method led to higher CTR values than two alternative approaches that do not consider user intents. Furthermore, the MIND method was deployed with a different number of latent intents, and the A/B test showed that choosing seven latent intents led to the best performance in terms of the CTR on this website. In~\cite{zhang_efficiently_2023}, Zhang et al.~report an +1.5\% increase in a ``\emph{top business metric}'' when deploying their \emph{Atten-Mixer} model on top of a tuned SR-GNN~\cite{SRGNN2019} model. These increases were observed after running the proposed model for a week on a heavy-traffic website with millions of page views per day. A Google research team reports the outcome of a three-week online test on a large scale website serving billions of users in~\cite{chang_latent_2023}. Two KPIs are considered in their study, \emph{enjoyment of the platform}\footnote{Details about how enjoyment was measured are not reported in the paper.} and topic-wise \emph{diversity of user-item interactions}. Comparing the proposed intent-aware sequential model with the previous intent-agnostic one led to a 0.7\% increase in enjoyment and a 0.1\% increase in diversity. Finally, in ~\cite{jannach_session-based_2017}, Jannach et al.~report the results of a field test in the e-commerce domain, where they examined the value of recommendations that serve the intent of \emph{reminding} users of items they had previously interacted with. A three-months A/B study on a e-commerce site for electronic gadgets revealed that reminding users of items they had previously viewed can be a highly effective strategy. The value-generating KPI in this study was the number of clicks on referral links to other online shops, which was substantially higher than when recommending the most popular items or when using BPR~\cite{rendle2009bpr} as an intent-agnostic recommendation model.

One single user study in form of an randomized controlled trial experiment was found in the surveyed papers.\footnote{The authors use the term ``field study'' for their experiment. The experiment was however not done in the form of a traditional A/B test in a production system.} In~\cite{yang_how_2019}, Yang et al.~modified an existing podcast app to include an onboarding phase where study participants could state their listening intents (interests). Furthermore, a collaborative filtering model was incorporated to populate the home feed, which was otherwise only filled with content from subscribed podcasts. In total, 99 participants concluded the study. During an onboarding phase, participants were tasked to provide their listening aspirations. The main study then lasted for four weeks, during which the participants' behavior was recorded. The experiment concluded with a post-study survey on their satisfaction. Among other findings, the obtained results indicated that intention-aware recommendations can help to significantly increase the participants' engagement in terms of listening behavior and subscriptions. Moreover, collaborative recommendations were found to be effective to increase the participants' interest in non-subscribed content and their overall satisfaction.

Finally, in addition to the work by Shi et al.~\cite{shi_improving_2017} discussed above, survey data was collected in ~\cite{mehrotra_jointly_2019} and~\cite{benedict_intent-satisfaction_2023}. The work by Mehrotra et al.~\cite{mehrotra_jointly_2019} reports on findings of a study at Spotify. A sample of three million Spotify users were presented with a survey in the app, where they were asked (a) about their satisfaction with their experience on the home screen, and (b) about the reasons of using the app on this day. For the latter question, six predefined options were presented which were identified through interviews with 12 Spotify users. Over 100.000 users responded to the survey, providing insightful information about the distribution of user intents in a real-world environment. Furthermore, the survey helped to identify user intents that were not considered in the preceding interview-based research. The entire research by Mehrotra et al.~was reproduced later in the video domain by Benedict et al.~\cite{benedict_intent-satisfaction_2023}. For their study, they also first interviewed user experience experts about possible user intents in this domain, and then conducted an in-app survey to obtain the distribution of intents and to learn about additional intents they had not thought of before.

\subsection{Evaluation Measures}
\begin{figure}[h!t]
    \centering
    \includegraphics[width=.5\textwidth]{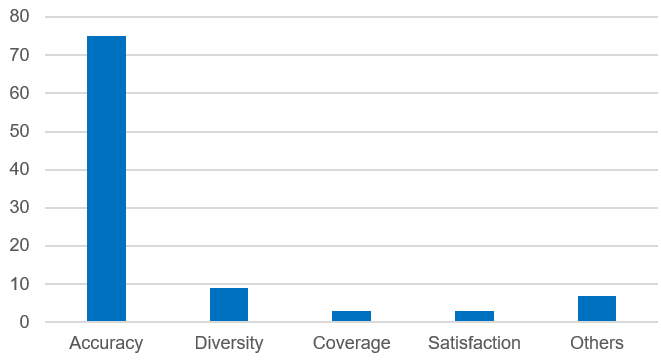}
    \caption{Metrics used in studied papers}
    \label{fig:metrics}
\end{figure}

Next, we analyzed which objectives researchers seek to optimize when proposing intent-aware recommendation approaches. Figure~\ref{fig:metrics} depicts the frequencies of metrics within the investigated set of papers. All frequencies of metrics occurring in less than three papers are aggregated in the category \emph{Others}.

Not too surprisingly, the majority of works focuses on improving the accuracy of the predictions (in offline experiments). This is expected, given that the main motivation of most IARS in the literature is to provide recommendations that better match the users' short-term intents. The second most frequent optimization goal is diversity, which is usually contrasted with a competing accuracy objective. Diversity is the `natural' optimization target for works that we classified as \emph{profile diversity matching} approaches in Section~\ref{sec:categorization}. Besides \emph{Catalog coverage} and \emph{Satisfaction}, the following additional, mostly business-oriented metrics were occasionally used in the literature:
unique visitors~\cite{yang_finn_2021,fan_metapath-guided_2019},
click-through-rates~\cite{yang_finn_2021,fan_metapath-guided_2019},
page views, clicks, add-to-cart events, and browsing time~\cite{shi_improving_2017,yang_finn_2021,fan_metapath-guided_2019},
enjoyment, user activity, and exploration behavior~\cite{yang_how_2019}, novelty~\cite{qian_intent_2022},
and entropy~\cite{Tomeo2015Exploiting}.

\subsection{Reproducibility}
\begin{figure}[h!t]
    \centering
    \includegraphics[width=.3\textwidth]{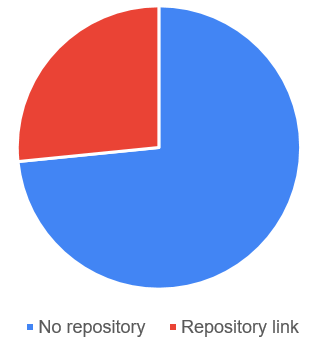}
    \caption{Percentage of papers with a repository link}
    \label{fig:reproducibility}
\end{figure}

Finally, we turn our attention to questions of reproducibility. Limited levels of reproducibility have shown to hamper progress not only in the field of recommender systems~\cite{dacremaetal2019,sun2020we}, but in the general field of AI~\cite{DBLP:conf/aaai/GundersenK18}. An in-depth study of the level of reproducibility of individual works is beyond the scope of this research. However, an analysis of the papers covered in this survey show that from the 75 papers that report offline experiments, only 21 (28\%) provide a link to a code repository, as visualized in Figure~\ref{fig:reproducibility}. This is a worrying observation, in particular given that previous reproducibility studies have indicated that sometimes complex neural architectures---and several modern IARS approaches have this characteristic---are not more effective than conceptual more simple models, see, e.g.,~\cite{shehzad2024performance}.

Considering other potentially problematic research practices, we found that in many surveyed papers that provide results of offline experiments, nothing is mentioned about the tuning process of the `state-of-the-art' baselines. Furthermore, in several works, the same embedding sizes are used across compared models, even though embedding sizes are a hyperparameter that has to be tuned for each model and dataset~\cite{shehzad2023everyone}.\footnote{We deliberately refrain from highlighting individual papers here.} Overall, these observations may lead to worries about the true progress that is achieved in the field of IARS in the last years. The research community therefore seems to be in need of in-depth reproducibility studies in this area.

\section{Discussion and Future Directions}
Our survey shows that researchers have become increasingly aware of the potential of intent-awareness as a powerful means to build the next generation of recommender systems. A variety of technical approaches have been proposed and successfully evaluated for different domains, both in offline experiments as well as in first field tests. In the following, we discuss open research gaps and possible directions for future works.

\subsection{Discussion and Gaps in Current Research}
\paragraph{Research Methodology} An apparent research gap, as discussed above, lies in the fact that today's research on intent-aware recommender systems is done almost entirely through offline experiments, often with limited reproducibility. Moreover, in these offline experiments we observed a strong focus on a single quality dimension, i.e., recommendation accuracy. It is therefore highly important that the research community more often conducts experiments with humans in the loop to truly understand the value of intent-awareness for users in different dimensions. On a positive note, our survey surfaced at least a few works which were validated in field tests with deployed systems. Unfortunately, several of the described A/B tests lasted only for a relatively short period of time, such as two or three weeks. Also, in several cases, central details such as the target KPI are not reported. We note that running an A/B test for three weeks may be considered sufficient in some settings, in particular when a large enough user base is involved in the experiment. However, papers often do not mention if there has been a ramp-up phase when a new model is deployed and do not report how the KPIs develop over time. If such factors are not considered, the long-term effects of a new model may easily be overestimated. For example, we may observe more interest (clicks) immediately after a  new model is deployed, when users explore the new types of recommendations. However, the more interesting and often unanswered question is if the new model truly has a lasting effect on user behavior. Generally, the limited number of field tests in the literature call for more intensive collaborations between academia and industry, both to evaluate new IARS models in a production environment and to create new real-world datasets for academic research.

\paragraph{Domains and Existing Datasets} In terms of application domains, the research scope on IARS is not different from the general literature on recommender systems, with a focus on the media and e-commerce domains. We also found that much of the literature is also based on well-known and widely used datasets. As these datasets do not contain explicit information about user intents, the user's intent has to be derived from the available information about users, items and user-item interactions. A certain gap therefore exists in terms of richer datasets, which either contain explicit information about intent\footnote{Explicit intent information could be acquired through in-app surveys, as in~\cite{mehrotra_jointly_2019}} or additional interaction signals that can be used for intent prediction. Interestingly, current research often does not leverage all the pieces of information that can be found in existing datasets, as mentioned above. Only a few works consider multiple types of behaviors in their models, e.g., clicks, purchases, or add-to-cart actions in e-commerce settings. Furthermore, some datasets like Diginetica\footnote{\url{https://competitions.codalab.org/competitions/11161}} contain rich information like user search queries. Intuitively, in particular search queries represent an important predictor for user intent, but current research does not yet exploit such information to a large extent. Another possible source of information for intent modeling and prediction is to rely on fine-grained logs of user interactions, such as clicks, mouse movements, or scrolling behavior. Such interaction logs were for example used in~\cite{bhattacharya_intent-aware_2017} and~\cite{sulikowski_fuzzy_2021}. Finally, Sun et al.~\cite{Lin2023Attention} demonstrated a novel way to leverage \emph{external} auxiliary information for intent modeling, in their case in the form of textual reviews.

Generally, both for already existing and future datasets, it is crucial to validate that these datasets are suited for the given task. In the area of sequential recommendation models, it was recently pointed out that much of the research done in the past few years is based on datasets that may actually not be truly sequential in nature~\cite{Woolridge2021Sequence,Hidasi2023widespread,klenitskiy2024Does}. Similar considerations apply for datasets used in IARS research, which often requires sequential data.

\paragraph{Explicit and Implicit Intents} Our survey showed that most current research works focuses on approaches that model intent in an implicit way or use latent intent models. A clear advantage of such approaches is that these models do not require application-specific knowledge, and that they can thus be easily applied in different application contexts. However, these models also come with certain disadvantages and limitations. First, the underlying intents remain latent, and the models therefore remain black boxes that are difficult to interpret. Our survey showed that the optimal number of latent intents in many approaches usually correspond with the number of manually identified intents in certain domains~\cite{mehrotra_jointly_2019,benedict_intent-satisfaction_2023}, e.g., between 4 and 16. This can be seen as a good sign. Nonetheless, we cannot be sure that the added model parameters truly represent user intent, and not just additional item co-occurrences or user interest trajectories in categories. Ultimately, the increased performance of complex IARS models might only stem from the fact that these models have more trainable parameters than then previous models they are compared with. We therefore sense a certain research gap in terms of works that explore models that rely on domain knowledge and explicit intents in a given domain. In sum, the strong focus on implicit intent modeling and the strong reliance of traditional datasets, as discussed above, may hamper progress towards novel or off-mainstream applications of recommender systems. Such novel applications could be represented by recommender systems that support users in their pursuit for self-actualization~\cite{Sullivan2019Reading,Knijnenburg2016Recommender} or healthier or more environmentally friendly behavior~\cite{trang_tran_overview_2018,Starke2021Using}.

On a more general level, a long-term ambition in algorithm-oriented research in recommender systems---and in other areas of applied machine learning---is to predict which of several models would perform best, given only the characteristics of the available data. Some work in that direction can for example be found in~\cite{adomavicius2012Impact}, but the problem in general seems largely unresolved. Translated to the domain of IARS, it would be similarly valuable to be able to know which approach---using implicit or explicit intents---should be adopted for a given use case by only considering the types of data and recorded interactions that are available. In the surveyed literature, we found no work that even tried to empirically compare the performance of explicit and implicit intent modeling approaches.

\subsection{Future Directions}

\paragraph{Leveraging Additional Information Sources}
The work by Rafique et al.~\cite{rafique_developing_2023} focuses on \emph{smart cities} application scenarios for IARS. The main assumption of their work is that in the future increasingly more traces of user behavior will be available, stemming from sensors, mobiles, Internet-of-Things devices etc., and that these pieces of information can be used for building next-generation recommender systems. In the current literature, sensor information is so far only used in a few selected works, e.g., in~\cite{lee_making_2014,changmai_-device_2019}. In addition to external sensor information, fine-grained user interaction data recorded by applications may also become frequently used in the future. Shi et al.~\cite{shi_improving_2017} and Sulikowski et al.~\cite{sulikowski_fuzzy_2021}, for example, used the user's mouse scroll speed as a predictor variable in their IARS. Besides more knowledge about user behavior, also more information about the users' situation and context, which may influence their intents, will become available through additional sensors or external knowledge sources. Finally, another direction lies in leveraging user-generated content such as text reviews provided by the user community, as proposed in~\cite{Lin2023Attention}.

Generally, research in IARS today is to some extent hampered by the limited availability of datasets that contain information or indications of possible user intents. Similar problems exist for other types of recommender systems, such as context-aware systems, price- and profit aware systems, or ones that consider multiple stakeholders~\cite{Adomavicius2015,debiasio2023economic,abdollahpouri2020}. However, this should not shy us away from directing our research efforts towards these highly relevant types of recommender systems. Instead, we believe that a paradigmatic shift is needed in how we do research on recommender systems. Instead of developing increasingly more complex models on a small set of frequently-used datasets that mainly consist of user-item interactions, an increased level of academia-industry collaborations is required. As mentioned above, this would allow us both to evaluate new models `in the wild' and to develop richer datasets for more realistic offline evaluations.

\paragraph{Better Understanding User Intents}
As indicated above, we see a strong potential in future works that aim at understanding the idiosyncrasies of individual application use cases, and explicitly considering potential user intents in a given setting. Ideally, such investigations should be supported either by an underlying theory, e.g., from psychology as in~\cite{he_intent-based_2014,wang_intention2basket_2020,wang_learning_2022}, or by empirical studies or domain-specific analyses, as done, e.g., for the music and video domains in~\cite{mehrotra_jointly_2019,benedict_intent-satisfaction_2023,Schaefer2013ThePsychological,schedl_current_2018,volokhin_understanding_2018}. A better understanding of these application-specific intents then serves as a basis for improved models for predicting the user's intent from their observed interactions, see, e.g.,~\cite{Cheng2017Predicting} using observational and survey data at Pinterest for improved intent prediction.

\paragraph{Interactive IARS}
We also envision important future directions in the context of the user experience of IARS. In practice, modern media streaming and e-commerce sites use multiple rows of recommendations to account for the various possible intents that users might have when arriving on the site. While multi-row user interfaces are common in practice, only a few works exist on this topic in academic research, e.g., \cite{PU2007542,jannachmultirow2021,DBLP:journals/tors/StarkeATL23}. More research is thus required in this area, for instance which intents should be supported in a given application setting, or how rows'content should be personalized and ordered; see~\cite{Gomez-Uribe:2015:NRS:2869770.2843948} for a discussion of the topic at Netflix. Furthermore, with the advent of LLM-based chatbots like ChatGPT, end users will become more and more be accustomed to natural language based advice-giving systems, and will increasingly use them as interactive and conversational recommender systems. In such usage scenarios, it may be quite common that end users will explicitly express their intents and situational context, e.g., ``\emph{I am planning to go for dinner with my friends after attending the Mets game on Sunday. Any recommendations?}'' A key challenge in future research will thus not only be to understand the expressed intent, to map it to suitable item recommendations, but even to provide a user interface that visualizes the assumed intent, see~\cite{Ruotsalo2015Interactive} for a related approach towards intent-modeling in interactive search.

\paragraph{Leveraging Large Language Models}
The emergence of LLMs is disrupting research in recommender systems, and a rich variety of proposals to leverage LLMs for recommendation were put forward since the release of the first models that were based on the Generative Pre-Trained Transformer (GPT) architecture. Common ways of incorporating LLMs into the recommendation process include the direct usage of an AI assistant like ChatGPT, various ways of model fine-tuning, generating prompts and prompt tuning, as well as using the `world knowledge' that is encoded in an LLM as a feature encoder, see~\cite{Wu2024ASurvey,Lin2024How,Zhao2024RSLLM}. The effectiveness of different approaches of incorporating LLMs to improve \emph{sequential} recommendation models was demonstrated in~\cite{Harte2023Leveraging}. Further, an iterative, prompt-based \emph{intent-aware} technique for session-based problems was recently proposed by Sun et al.~\cite{Sun2024Large}, leading to improved offline accuracy results. A notable aspect when using LLMs for intent-aware recommendation with the help of prompts is that an LLM can provide explanations and its chain-of-thought regarding the assumed current user intent. Given their rich knowledge, LLMs may also serve as a powerful tool to list possible user intents for a given recommender systems use case, thereby laying the foundation for future Explicit Intent Modeling approaches.

\section{Summary}
We argue that intent-awareness can be a key building block for the next generation of recommender systems. With this paper, our aim is to provide an overview of the existing literature on IARS, which shall serve as a starting point for researchers working in this area. We have therefore searched the literature for relevant papers and proposed a categorization into different dimensions. Our analyses points to a number of open research gaps, for instance related to the so far limited evaluation of current IARS with humans in the loop. Finally, we outlined a number of possible research directions, which include a stronger focus on domain specifics and the use of additional types of information to infer the current intent of users.

\bibliographystyle{ACM-Reference-Format}


\end{document}